\documentstyle[epsfig,sprocl]{article}

\def\Journal#1#2#3#4{{#1} {\bf #2}, #3 (#4)}

\def\NPA{{\em Nucl. Phys.} A}

\def\PLB{{\em Phys. Lett.}  B}
\def\PRL{\em Phys. Rev. Lett.}

\def\PRC{{\em Phys. Rev.} C}

\def\be{\begin{equation}}
\def\ee{\end{equation}}
\def\bea{\begin{eqnarray}}
\def\eea{\end{eqnarray}}

\begin{document}

\title{SEARCHING FOR SPACE-TIME ASYMMETRIES IN PARTICLE PRODUCTION}

\author{R. LEDNICK\'Y}

\address{Institute of Physics, Na Slovance 2, 18040 Prague 8,
Czech Republic and SUBATECH, 4, rue Alfred Kastler,
F-44070 Nantes Cedex 03, France
\\E-mail: lednicky@fzu.cz}


\maketitle\abstracts{The possibilities of unlike particle
correlations for a study of the space-time asymmetries in particle
production, including the sequence of particle emission, 
are demonstrated.}

\section{Introduction}
The correlations of particles at small relative velocities
are widely used to study space-time characteristics of the
production processes.
Particularly, for non-interacting identical particles, like photons, 
this technique is called intensity or particle interferometry.
In this case the correlations appear solely due to the effect of 
quantum statistics (QS).\cite{GGLP60,KP72}
Similar effect was first used 
in astronomy to measure
the angular radii of stars by studying the dependence of the two-photon
coincidence rate on the distance between the detectors
(HBT effect \cite{hbt}).
In particle physics the QS interference was first
observed as an enhanced production of the
pairs of identical pions with small opening angles 
(GGLP effect \cite{GGLP60}). Later on, similar to astronomy,
Kopylov and Podgoretsky \cite{KP72}
suggested to study the interference effect in terms of the correlation
function.\footnote
{Note
that though both the KP and HBT methods are based on 
the QS interference, they represent just orthogonal measurements.\cite{KP72}
The former, being the momentum-energy measurement, yields the
space-time picture of the source, while the latter does the opposite.
In particular, the HBT method provides the information about the angular
size of a star  
but, of course, - no information about the star radius or its
lifetime. 
}  

The effect of QS is usually considered in the limit of a low
phase-space density such that the possible multi-particle effects can be
neglected.
This approximation seems to be justified by present experimental
data which does not point to any spectacular multi-boson effects neither
in single-boson spectra nor in two-boson correlations.
These effects can be however of some importance for realistic simulations
of heavy ion collisions.\cite{ame-led}
They can also clearly manifest themselves in some rare events
({\it e.g.}, those with large pion multiplicities) or in
the eventually overpopulated regions of momentum
space (see, {\it e.g.},\cite{pra93,ALICE} 
and references therein).\footnote
{ 
It was shown,\cite{pra93,ALICE} using the analytically solvable 
model with the Gaussian distribution of 
the emission points characterized by the dispersions $\Delta^2$ 
and $r_0^2$ in the momentum and ordinary space respectively,
that a pion condensate of
a small momentum dispersion of \cite{ALICE}
$\Delta/2r_0 \ll \Delta^2$ gradually develops
with the increasing phase-space density. }

The particle correlations are also
influenced by the effect of particle interaction 
in the final state (FSI).\cite{koo,ll1}
Thus the effect of the Coulomb interaction dominates the correlations
of charged particles at very small relative momenta
(of the order of the inverse Bohr radius of the two-particle system), 
respectively suppressing or 
enhancing the production of particles with like or unlike charges.
Regarding the effect of the strong FSI, it is 
quite small for pions, while for nucleons it is often a dominant one
due to the very large magnitude of the s-wave singlet scattering length
of about 20 fm.

Though the FSI effect complicates the correlation analysis,
it is an important source of information allowing one to measure
the space-time characteristics of the production process even with the
help of non-identical particles.\cite{ll1,BS86} 
Moreover, the unlike particle correlations, 
contrary to those of identical particles, 
are sensitive to the relative space-time asymmetries in their production,
{\it e.g.} - to the relative time delays,
thus giving an important complementary information not accessible
to the standard interferometry measurements.\cite{LLEN95}
In the following we will briefly formulate the theory of these correlations
and demonstrate the possibilities of the corresponding correlation technique.

\section{Formalism}\label{sec:Formalism}
As usual, we will assume
sufficiently small phase-space
density of the produced multi-particle system,
such that the correlation of two particles
emitted with a small relative
velocity in nearby space-time points is influenced by the effects of
their mutual QS and FSI only.\footnote
{
This assumption may be not valid in the case of low energy heavy ion reactions
when the particles are produced in a strong Coulomb field of residual
nuclei. To deal with this field a quantum $adiabatic$
approach can be used.\cite{LLEN95}
} 
We define the ideal two-particle
correlation function $R(p_{1},p_{2})$ as a
ratio of the differential two-particle production cross section to the
reference one which would be observed in the absence of the
effects of QS and FSI. 
In heavy ion  or high energy hadronic collisions 
we can neglect kinematic constraints
and most of the dynamical correlations and construct the
reference distribution by mixing the particles from different
events. 

Assuming the momentum dependence of the one-particle emission probabilities
inessential when varying the particle 4-momenta $p_{1}$ and $p_{2}$ by the
amount characteristic for the correlation due to QS and FSI
({\it smoothness assumption}),
{\it i.e.} assuming that
the components of the mean space-time distance
between particle emitters 
are much larger than those of the space-time extent
of the emitters, 
we get the well-known result of Kopylov and Podgoretsky for
identical particles,
modified by the substitution of the plane wave
$e^{ip_{1}x_{1}+ip_{2}x_{2}}$ by the nonsymmetrized Bethe-Salpeter
amplitudes in the continuous spectrum of the two-particle states
$\psi_{p_{1}p_{2}}^{S(+)}(x_{1},x_{2})$,
where $x_{i}=\{t_{i},{\bf r}_{i}\}$ are the 4-coordinates 
of the emission points
of the two particles and $S$ is their total spin.\cite{ll1}
At equal emission times in the
two-particle c.m.s. ($t^{*}=t_{1}^{*}-t_{2}^{*}=0$) this amplitude
coincides (up to an unimportant phase factor due to the c.m.s. motion)
with a stationary solution  of the scattering problem
$\psi_{-{\bf k}^{*}}^{S(+)}({\bf r}^{*})$,
where 
${\bf k}^{*}= {\bf p}^{*}_{1} = -{\bf p}^{*}_{2}$
and ${\bf r}^{*}= {\bf r}^{*}_{1} -{\bf r}^{*}_{2}$
(the minus sign of the vector ${\bf k}^{*}$ corresponds to the reverse
in time direction of the emission process). 
The Bethe-Salpeter amplitude can be usually substituted  by
this solution ({\it equal time} approximation).\footnote
{The {\it equal time} approximation is valid
on condition \cite{ll1}
$ |t^*|\ll m_{2,1}r^{*2}$ for 
${\rm sign}(t^*)=\pm 1$ respectively.
This condition is usually satisfied
for heavy particles like kaons or
nucleons. But even for pions, the $t^{*}=0$ approximation
merely leads to a slight overestimation (typically $<5\%$) of the strong
FSI effect and, 
it doesn't influence the leading zero--distance 
($r^{*}\ll |a|$) effect of the Coulomb FSI.}
Then, for nonidentical particles,
\begin{equation}
R(p_{1},p_{2})=
\sum_{S}\rho_{S}
\langle |\psi_{-{\bf k}^{*}}^{S(+)}({\bf r}^{*})|^{2}
\rangle _{S}.
\label{1}
\end{equation}
Here the averaging is done over the emission points
of the two particles in a state with total spin $S$
populated with the probability
$\rho_{S}$,
$\sum_{S}\rho_{S} = 1$.\footnote
{
For unpolarized particles
with spins $s_{1}$ and $s_{2}$ the probability 
$\rho_{S}=(2S+1)/[(2s_{1}+1)(2s_{2}+1)]$.
Generally, the correlation function is sensitive to particle
polarization. For example, if two spin-1/2 particles are emitted with 
polarizations ${\bf P}_1$ and ${\bf P}_2$ then \cite{ll1}
$\rho_0=(1-{\bf P}_1\cdot{\bf P}_2)/4$ and
$\rho_1=(3+{\bf P}_1\cdot{\bf P}_2)/4$.
}
For identical particles, the amplitude in Eq.~(\ref{1})
should be properly symmetrized:
\begin{equation}
\label{3}
\psi_{-{\bf k}^{*}}^{S(+)}({\bf r}^{*}) \rightarrow
[\psi_{-{\bf k}^{*}}^{S(+)}({\bf r}^{*})+(-1)^{S}
\psi_{{\bf k}^{*}}^{S(+)}({\bf r}^{*})]/\sqrt{2}.
\end{equation}

\section{Measuring the relative space-time asymmetries}

The correlation function of two nonidentical particles, compared with the
identical ones,
contains a principally new piece of information on the relative
space-time asymmetries in particle emission.\cite{LLEN95}
This is clearly seen in the case of neutral particles
when the two-particle amplitude
$\psi_{-{\bf k}^{*}}^{S(+)}({\bf r}^{*})$
takes on the form
\begin{equation}
\label{6}
\psi_{-{\bf k}^{*}}^{S(+)}({\bf r}^{*}) =
e^{-i{\bf k}^{*}{\bf r}^{*}}+\phi^{S}_{k^{*}}(r^{*}),
\end{equation}
where the scattered wave $\phi^{S}_{k^{*}}(r^{*})$, in the considered
region of small relative momenta, is independent of the
directions of the vectors ${\bf k}^{*}$ and ${\bf r}^{*}$.
Inserting Eq. (\ref{6}) into the formula (\ref{1}) for 
the correlation function, we can see that the latter 
is sensitive to the relative space-time asymmetry due to the
odd term $\sim\sin {\bf k}^{*}{\bf r}^{*}$. 
Particularly, it allows for a measurement of the mean relative delays
$\langle t\rangle\equiv\langle t_1-t_2\rangle$ in particle emission.
To see this, let us make the Lorentz
transformation from the rest frame of the source
to the c.m.s. of the
two particles:
$r_{L}^{*} = \gamma (r_{L}-vt)$, $r_{T}^{*}=r_{T}$.
Considering, for simplicity, the behavior of the vector ${\bf r}^{*}$
in the limit $|vt| \gg r$, 
we see that this vector
is only slightly affected by averaging over the spatial
distance $r \ll |vt|$ of the emission points in the rest frame
of the source so that 
${\bf r}^{*}\approx -\gamma{\bf v}t$.
Therefore, the vector ${\bf r}^{*}$
is nearly parallel or antiparallel to the velocity vector
${\bf v}$ of the pair, depending on the sign of the time
difference $t\equiv\Delta t=t_1-t_2$.
The sensitivity to this sign is transferred to the correlation
function through the odd in
${\bf k}^{*}{\bf r}^{*}\approx  -\gamma{\bf k}^{*}{\bf v}t$ term 
provided the sign of the scalar product
${\bf k}^{*}{\bf v}$ is fixed.

For charged particles there arise additional odd terms due to
the confluent hypergeometrical function 
$F(\alpha,1,z)=1+\alpha z/1!^2+\alpha(\alpha+1)z^2/2!^2+\dots$, 
modifying the plane wave in Eq.~(\ref{6}):
\begin{equation}
\psi_{-{\bf k}^{*}}^{S(+)}({\bf r}^{*}) = {\rm e}^{i\delta}
\sqrt{A_{c}(\eta)}\left[
{\rm e}^{-i{\bf k}^{*}{\bf r}^{*}}F\left(-i\eta,1,i\rho\right)
+\phi^{S}_{ck^{*}}(r^{*})\right],
\label{14}
\end{equation}
where 
$\rho={\bf k}^*{\bf r}^*+{\rm k}^*r^*$, $\eta=(k^{*}a)^{-1}$,
$a=(\mu z_{1}z_{2}e^{2})^{-1}$ is the Bohr radius of the
two-particle system taking into account the sign of the interaction
($z_ie$ are the particle electric charges,
$\mu$ is their reduced mass),
$\delta=\mbox{arg}\Gamma(1+i\eta)$
is the Coulomb s-wave shift and
$A_{c}(\eta)=2\pi\eta/[\exp (2\pi\eta)-1]$
is the Coulomb penetration factor.\footnote
{This factor
substantially deviates from unity
only at $k^{*}< 2\pi/|a|$ ({\it e.g.}, at
$k^{*} < 22$ MeV/c for two protons).
Note that for the distances $r^*>|a|$ the 
confluent hypergeometrical function becomes important and
compensates the deviation of the Coulomb factor from unity
except for the classically
forbidden region of $k^{*} <  (|a|r^{*}/2)^{-1/2}$,
narrowing with the increasing $r^*$.
Thus the FSI practically vanishes if at least one of the two particles 
comes from a long lived
source ($\eta, \eta', \Lambda, K^0_s, \ldots$).
}
Clearly, at a given distance $r^*$, the effect of the odd component
in the Coulomb wave function is of increasing importance with
a decreasing Bohr radius of the particle pair,
{\it i.e.} for particles
of greater masses or electric charges.
At low energies,
the sensitivity of the correlation to the odd component can be somewhat
modified due to Coulomb interaction with the residual charge.\cite{LLEN95}
 
It is clear that in the case of a dominant time asymmetry, 
$v|\langle t\rangle| \gg |\langle r_L\rangle|$,
a straightforward way to determine
the mean time delay $\langle t\rangle $ is to
measure the correlation functions
$R_{+}({\bf k}^{*}{\bf v}\geq 0)$ and
$R_{-}({\bf k}^{*}{\bf v}< 0)$.
Depending on the sign of $\langle t\rangle$,
their ratio $R_{+}/R_{-}$ should show a peak
or a dip in the region of small $k^{*}$ and approach 1
at large values of $k^{*}$.
As the sign of the scalar product
${\bf k}^{*}{\bf v}$ is practically equal to that of the
difference of particle velocities $v_{1}-v_{2}$
(this equality is always valid for particles of equal masses),
the sensitivity of the correlation functions
$R_{+}$ and $R_{-}$ to the sign of the difference of particle
emission times has a simple classical explanation. Clearly,
the interaction between the particles in the case of an earlier
emission of the faster particle will be weaker compared
with the case of its later emission (the interaction time being
longer in the latter case leading to a stronger correlation).
This expectation is in accordance with 
Eqs. (\ref{1}) and (\ref{14})
at $k^* \rightarrow 0$, $\langle r^*\rangle \ll |a|$ and
$\langle |\phi^{S}_{ck^{*}}(r^{*})|\rangle \ll 1$, when
(the arrow indicates the limit
$v|\langle t\rangle| \gg |\langle r_L\rangle|$):
\begin{equation}
R_+/R_-\approx 1+
2\frac{\langle r_L^*\rangle }{a} \rightarrow
1-2\frac{\langle \gamma v(t_1-t_2)\rangle }{a}.
\label{51}
\end{equation}
 
The sensitivity of the $R_+/R_-$ correlation method to the mean
relative time shifts (introduced {\it ad hoc}) 
was studied\cite{ALICE}
for various two-particle systems 
simulated in $Pb+Pb$ collisions at SPS energy
using the event generator VENUS 5.14.\cite{WER93}
The scaling of the effect with the space-time asymmetry and 
with the inverse Bohr radius $a$,
indicated by Eq. (\ref{51}), was clearly illustrated
for the $K^+K^-$ system ($a=-110$ fm)
and for the like- and unlike-sign $\pi K$, $\pi p$ and $Kp$
systems ($a=\pm 249$, $\pm 226$ and $\pm 84$ fm respectively).                                                                  
It was concluded that for sufficiently relativistic pairs
($\gamma v > 0.5$)
the $R_{+}/R_{-}$ ratio can be sensitive to the shifts in the
particle emission times of the order of a few fm/c. 
Motivated by this result the $R_+/R_-$ method was  recently 
applied to the $K^+K^-$ system
simulated in a two-phase thermodynamical evolution model
and the sensitivity was demonstrated
to the production of the transient strange quark matter state
even if it decays on strong interaction time scales.\cite{sof97}

\begin{figure}[hb] 
\begin{center}\mbox{ 
                \epsfxsize=10cm
                 \epsffile{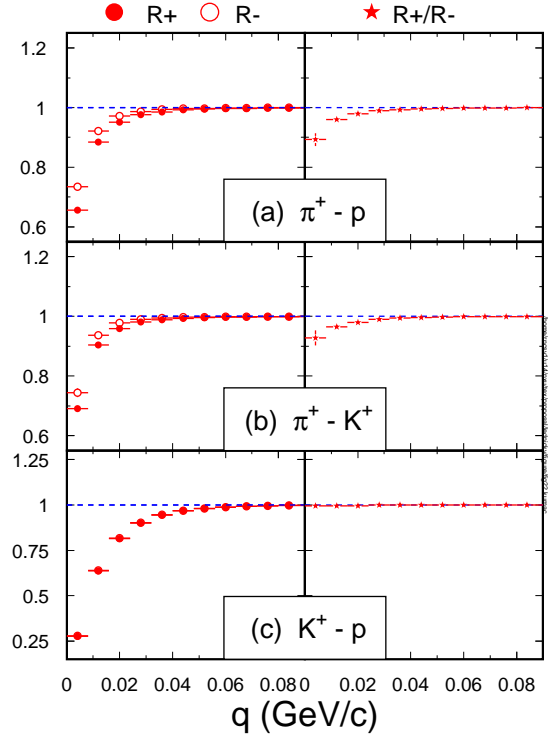}
}
\end{center}
\caption{ 
Unlike particle correlation functions 
$R_+$ and $R_-$ and their ratios simulated 
with RQMD
for mid-rapidity particle
pairs $\pi^+p$, $\pi^+K^+$, and $K^+p$ 
in $Pb+Pb$ collisions at SPS energy.
}
\label{pikpSPS}
\end{figure}

The method sensitivity was also studied 
for AGS and SPS energies using the transport code RQMD v2.3.\cite{rqmd}
Thus at SPS energy the central $Pb+Pb$ collisions have been simulated
and the $\pi^+K^+$, $\pi^+p$ and $K^+p$ correlations 
have been studied.\cite{lpx} 
To get rid of the effect of a fast longitudinal motion,
the study was done in the longitudinally co-moving system (LCMS)
in which the pair is emitted transverse
to the reaction axis so that $v=v_{\perp}$, $r_L=\Delta x$ and
\begin{equation}
\label{delta}
\Delta x^*=\gamma_{\perp}(\Delta x-v_{\perp}\Delta t),~~\Delta y^*=\Delta y,~~
\Delta z^*=\Delta z.
\end{equation}
The simulated correlation functions $R_+$, $R_-$ and their ratios are plotted 
in Fig. \ref{pikpSPS}.
We can see that for $\pi ^+p$ and $\pi ^+K^+$ systems these ratios are less than
unity at small values of $q\equiv k^*$, 
while for $K^+p$ system the ratio $R_+/R_-$
practically coincides with unity. These results well agree with the mean values
of $\Delta t$,  $\Delta x$ and $\Delta x^*$ presented in Table 1
($\langle\Delta y\rangle\approx\langle\Delta z\rangle\approx 0$).
It can be seen from Eqs. (\ref{51}) and (\ref{delta}) 
that the absence of the effect in the $R_+/R_-$ ratio for the  $K^+p$
system is due to practically the complete compensation of the space and time
asymmetries leading to $\Delta x^* \approx 0$. For $\pi^+p$ system the effect
is determined mainly by the x-asymmetry. For $\pi^+K^+$ system both the
x- and time-asymmetries contribute in the same direction, 
the latter contribution
being somewhat larger.
The separation of the relative time delays from the
spatial asymmetry is, in principle, possible (see Eq. (\ref{delta}))
by studying the ratio $R_+/R_-$ in different intervals of the
pair velocity.

At AGS energy 
the $Au+Au$ collisions have been simulated and 
the $\pi^+p$ correlations have been studied
in the projectile fragmentation region where proton directed
flow is most pronounced and where the proton and pion sources
are expected to be shifted relative to each other both
in the longitudinal
and in the transverse (flow) directions in the reaction plane.
It was demonstrated that a modification
of the $R_{+}/R_{-}$ method ($\pm$ corresponding now to the signs of the
respective components $k^*_i$) is sufficiently sensitive to reveal these
shifts.\cite{vol97}

At low energies, the particles in heavy ion collisions are emitted
with the characteristic emission times of tens to hundreds fm/c so that
the observable time shifts should be of the same order.\cite{LLEN95}
In fact the $R_{+}/R_{-}$ method has already been successfully applied
to study proton-deuteron correlations in several heavy ion
experiments at GANIL.\cite{ghi95,nou96,gou96} 
It was observed, in agreement 
with the coalescence model, that deuterons are on average emitted earlier
than protons.

\section{Conclusion}

We have shown that unlike particle
correlations, compared with those of
identical particles,
contain a principally new piece of information on the relative
space-time asymmetries in particle emission, thus allowing,
in particular, a measurement of the mean relative delays
in particle emission at time scales as small as $10^{-23}$ s. 
To determine these asymmetries, the unlike particle
correlation functions $R_+$ and $R_-$ have to be studied
separately for positive and negative values of the projection
of the relative momentum vector in pair c.m.s. on the 
pair velocity vector or, generally, - on any direction of interest.
We have presented here the results of recent studies of these
correlation functions for
a number of two-particle systems simulated with various event generators.
It was shown
that the $R_+/R_-$ ratio is sufficiently sensitive to the
relative space-time asymmetries arising
due to the formation of the quark-gluon plasma and strangeness
distillation and even to those expected in the usual dynamical scenarios
at AGS and SPS energies. 
As to the detection of the unlike particles with close velocities 
($p_1/m_1\approx p_2/m_2$),
there is no problem with the two-track resolution since these
particles,
having either different momenta or different charge-to mass ratios,
have well separated trajectories in the detector magnetic field.
For the same reason, however, a large momentum acceptance of the
detector is required.

\section*{Acknowledgments}
This work was supported by GA AV Czech Republic, Grant No. A1010601
and by GA Czech Republic, Grant No. 202/98/1283.


\begin{table}[htbp]
\caption[]{\footnotesize
Mean values of the relative space-time coordinates in LCMS
(in fm) calculated \cite{lpx} from
RQMD (v2.3) for 158 A$\cdot$GeV $Pb+Pb$ central collisions.
}
\label{trq}

\medskip
\begin{center}
\begin{tabular}{|c|c|c|c|c|c|} 
\hline
     &system & $\langle\Delta t\rangle$ & $\langle\Delta x\rangle$
     & $\langle\Delta x -v_{\perp}\Delta t\rangle$
     & $\langle\Delta x^*\rangle$\\
\hline
        &$\pi^+p$   &-0.5 &-6.2 &-6.4 &-7.9\\
        &$\pi^+K^+$ & 4.8 &-2.7 &-5.8 &-7.9\\  
        &$K^+p$    &-5.5 &-3.2 &-0.6 &-0.4\\
\hline
\end{tabular}
\end{center}
\end{table}

\section*{References}
\begin{thebibliography}{99}


\bibitem{GGLP60}
G.~Goldhaber et al., {\it Phys.~Rev.} {\bf 120}, 300 (1960).

\bibitem{KP72}
G.I.~Kopylov, M.I.~Podgoretsky, {\it Yad.~Fiz.} {\bf 15}, 392 (1972)
({\it Sov. J. Nucl. Phys.} {\bf 15}, 219 (1972));
G.I.~Kopylov, \Journal{\PLB}{50}{472}{1974};
M.I.~Podgoretsky,
{\it Fiz.~Elem. Chast. Atom. Yad.} {\bf 20}, 628 (1989)
({\it Sov. J. Part. Nucl.} {\bf 20}, 266 (1989)).

\bibitem{hbt}
R.~Hanbury-Brown, R.Q.~Twiss,
{\it Phil. Mag.} {\bf 45}, 663 (1954);
{\it Nature} {\bf 178}, 1046 (1956).

\bibitem{ame-led}
N.S.~Amelin, R.~Lednicky,
{\it Heavy Ion Physics} {\bf 4}, 241 (1996);
SUBATECH 95-08, Nantes 1995.

\bibitem{pra93}
S.~Pratt,
\Journal{\PLB}{301}{159}{1993};
\Journal{\PRC}{50}{469}{1994}.

\bibitem{ALICE} B. Erazmus et al., Internal Note ALICE 95-43.

\bibitem{koo}
S.E.~Koonin, \Journal{\PLB}{70}{43}{1977}; 
M.~Gyulassy, S.K.~Kauffmann, L.W.~Wilson,
\Journal{\PRC}{20}{2267}{1979}.

\bibitem{ll1}
R.~Lednicky, V.L.~Lyuboshitz,
{\it Yad.~Fiz.} {\bf 35}, 1316 (1982) ({\it Sov. J. Nucl. Phys.} 
{\bf 35}, 770 (1982));
Proc. CORINNE 90,
Nantes, France, 1990 (ed. D.~Ardouin, World Scientific, 1990) p. 42;
JINR report P2-546-92 (1992); 
{\it Heavy Ion Physics} {\bf 3}, 93 (1996).

\bibitem{BS86}
D.H.~Boal, J.C.~Shillcock, 
\Journal{\PRC}{33}{549}{1986};
D.H.~Boal, C.-K. Gelbke, B.K.~Jennings,
{\it Rev. Mod. Phys.} {\bf 62}, 553 (1990).

\bibitem{LLEN95}
R.~Lednicky, V.L.~Lyuboshitz, B.~Erazmus, D.~Nouais,
\Journal{\PLB}{373}{30}{1996};
Report SUBATECH 94-22, Nantes 1994.


\bibitem{WER93} K.~Werner,
{\it Phys.~Rep.} {\bf 232}, 87 (1993);  
K. Werner and J. Aichelin,
\Journal{\PRC}{52}{1584}{1995}. 

\bibitem{sof97} S.~Soff et al.,
{\it J.~Phys.} {\bf G12}, 2095 (1997).

\bibitem{rqmd} H.~Sorge et al.,
\Journal{\PLB}{243}{7}{1990}.

\bibitem{lpx} R.~Lednicky, S.~Panitkin, Nu Xu,
submitted to QM97, Tsukuba.

\bibitem{vol97} S.~Voloshin, R.~Lednicky, S.~Panitkin, Nu Xu,
\Journal{\PRL}{79}{4766}{1997}.

\bibitem{ghi95} C.~Ghisalberti et al.,
\Journal{\NPA}{583}{401}{1995}.

\bibitem{nou96} B. Erazmus et al., Proc. XXXIV Bormio Meeting (1996);
D.~Nouais, PhD Thesis (1996), Nantes (unpublished).

\bibitem{gou96} D.~Gourio, PhD Thesis (1996), Nantes (unpublished).

\end{thebibliography}
\end{document}